\documentclass[11pt,fleqn]{article}
\usepackage{amsmath,amssymb,graphicx,epsfig,latexsym}
\numberwithin{equation}{section}
\usepackage{color}
\usepackage{hyperref}
\def\bea{\begin{eqnarray}}
\def\eea{\end{eqnarray}}
\def\be{\begin{equation}}
\def\ee{\end{equation}}
\def\ba{\begin{array}}
\def\ea{\end{array}}

\font\tenrsfs=rsfs10
\font\sevenrsfs=rsfs7
\font\fiversfs=rsfs5
\newfam\rsfsfam
\textfont\rsfsfam=\tenrsfs
\scriptfont\rsfsfam=\sevenrsfs
\scriptscriptfont\rsfsfam=\fiversfs
\def\mathscr#1{{\fam\rsfsfam\relax#1}}

\begin{document}

\thispagestyle{empty}
\begin{flushright}
CERN-PH-TH/2012-084 \\
April 2012
\end{flushright}
\vspace{1.0cm}

\begin{center}

$\;$

\vspace{1cm}

{\huge \bf $\mathbf{\mathcal N=2}$ supersymmetry breaking at two different scales}

\vspace{1.3cm}

{\Large {\bf Ignatios Antoniadis$^{1 \dagger}$
, Jean-Pierre Derendinger$^{2}$} and {\bf Jean-Claude Jacot$^{1,2}$}}\\[2mm] 

\vspace{0.4cm}

{\large $^1$ \em Department of Physics, CERN Theory Division\\ 
CH-1211 Geneva 23, Switzerland\\ }

\vspace{0.4cm}

{\large $^2$ \em Albert Einstein Center for Fundamental Physics \\
Institute for Theoretical Physics, Bern University \\
Sidlerstrasse 5, CH--3012 Bern, Switzerland \\}

\vspace{0.2cm}

\end{center}

\vspace{1cm}

\centerline{\bf \large Abstract}
\begin{quote}
We study ${\mathcal N}=2$ spontaneous supersymmetry breaking at two different scales with matter fields in hypermultiplets charged under the gauge group that should involve at least two $U(1)$ factors. Off-shell analysis is possible in the dual single-tensor formulation of the hypermultiplets.  Massless fermions can naturally arise from pseudo-real representations of the gauge group that allow a reformulation of the problem of chirality in ${\mathcal N}=2$ theories. The above properties are necessary ingredients towards constructing viable extensions of the Standard Model based on ${\mathcal N}=2$ supersymmetry.
\vspace{5cm}

{$^\dagger$\small\em On leave from CPHT (UMR CNRS 7644) Ecole Polytechnique, F-91128 Palaiseau\\ }

\end{quote}

\renewcommand{\theequation}{\thesection.\arabic{equation}}


\setcounter{page}{1}
\section{Introduction}
Spontaneous breaking of $\mathcal{N} =2$ supersymmetry (SUSY) is an old subject \cite{Fayet:1975yi}, but it has been comparatively less studied than the breaking of $\mathcal N=1$ SUSY. One reason for this disparity is that the breaking of $\mathcal N=2$ SUSY is quite constrained (with $\mathcal N=2$ Fayet-Iliopoulos (FI) terms playing a central role), while $\mathcal N=1$ SUSY breaking can occur in many different ways.

An important additional avenue for $\mathcal N=2$ SUSY breaking was discovered in \cite{Antoniadis:1995vb}. Inspired by work on electric-magnetic duality \cite{Seiberg:1994rs}, the authors of \cite{Antoniadis:1995vb} realized that one can deform an $\mathcal N=2$ abelian gauge theory by a superpotential term linear in the first derivative of the prepotential. Such a deformation is referred to as a magnetic FI term, and the resulting theory is invariant under a deformed $\mathcal N=2$ SUSY (current) algebra that allows partial breaking of $\mathcal N=2$ to $\mathcal N=1$ (thus evading a famous no-go theorem which applies only to the undeformed algebra).

There are many reasons to think that the short-distance theory describing our world may contain some sectors that are invariant under extended SUSY. For example, theories with extended SUSY naturally appear when one considers stacks of D-branes wrapping cycles in certain internal geometries. From a more bottom-up perspective, theories with gauge subsectors that have a remnant of $\mathcal N=2$ supersymmetry give rise to interesting phenomenological signatures \cite{Fox:2002bu} and may be important for describing new physics at the LHC (see e.g. \cite{Antoniadis:2006uj, Benakli:2010gi, Choi:2010gc, Heikinheimo:2011fk} and references therein). 

In spite of this motivation, one major stumbling block for imagining that $\mathcal N=2$ plays an important role in short-distance physics is that the standard matter representations of $\mathcal N=2$ SUSY are non-chiral. Therefore, all the Standard Model (SM) matter particles would necessarily have mirror particles with the opposite internal quantum numbers. In order to try to overcome this obstacle, one could imagine constructing theories that spontaneously break the two supersymmetries at two widely separated scales and then hope that chirality could somehow be recovered in the low-energy theory.\footnote{One intriguing idea along these lines would be for the effective $\mathcal N=1$ theory below the first SUSY breaking scale and above the second SUSY breaking scale to undergo a non-chiral / chiral duality similar to the type discussed in \cite{Pouliot:1995zc}.} Another interesting possibility is to use pseudo-real representations to construct hypermultiplets for the UV ancestors of phenomenologically-relevant matter fields that are naturally light, and then find a mechanism forcing heavy mirrors.

In this paper, we will focus mainly on constructing relatively simple models that break the two SUSYs at parametrically different scales (to our knowledge, the models we will describe below are the first models that break SUSY at two different scales). As we will see below, magnetic FI terms will play a crucial role in allowing this to happen. While our simple examples do not have a chiral spectrum in the low energy limit, we will also briefly comment on the possibility of using pseudo-real representations to construct theories with chiral spectra. We hope to use these tools and return to the question of whether or not one can reproduce the chiral spectrum of the SM in a future publication.\footnote{Previous attempts to build viable $\mathcal N=2$ supersymmetric extensions of the Standard Model are for example \cite{delAguila:1984qs,Girardello:1997hf,Polonsky:2000zt}. Their approach to the problem of chirality is briefly reviewed in section \ref{chir}.}

The paper is organized as follows: in section \ref{mod}, we present our (minimal) model based on a $U(1)\times U(1)$ gauge group in detail. Using a single-tensor multiplet to describe the addition of some bifundamental matter fields, we argue that introducing the second $U(1)$ gauge factor is necessary for our gauge theory with matter to be invariant under the deformed supersymmetry algebra of \cite{Antoniadis:1995vb}. In section \ref{vac}, we investigate the properties of the vacua in the Higgs and Coulomb phases. After having discussed the structure of vacua, section \ref{gold} focuses on supersymmetry breaking. We will use expressions for the two potential goldstini in order to characterize partial SUSY breaking, and, especially, to show that the presence of magnetic FI terms allows the two supersymmetries to be broken at two different energies. In section \ref{chir}, we will comment on the possibility of obtaining naturally light fermionic states in $\mathcal N=2$ theories and its role in the problem of chirality. We conclude in section \ref{conc}. Throughout this note, we use the conventions of \cite{Wess:1992cp}.

\section{Review of the model in $\mathcal N=1$ and $\mathcal N=2$ superspace\label{mod}}
Firstly we present the $\mathcal N=2$ supersymmetric model of interest in terms of $\mathcal N=2$ superfields in $\mathcal N=2$ superspace, where invariance under a second supersymmetry is manifest. Then we derive the more familiar expression of the action in $\mathcal N=1$ superfields.

The description of $\mathcal N=2$ gauge multiplets in ordinary $\mathcal N=2$ superspace is well-known.\footnote{See for example \cite{Ambrosetti:2009za} for a recent review.} However, it is also well established that an equivalent off-shell expression for hypermultiplets requires for instance the more involved harmonic superspace \cite{Galperin:2001uw}. In the presence of abelian isometries, it is nevertheless possible to use the ordinary superspace for the single-tensor multiplet, which is dual to the hypermultiplet \cite{Lindstrom:1983rt}.
\subsection{Maxwell multiplet}
We review our conventions for $\mathcal N=2$ superspace with the presentation of the gauge multiplet. We shall restrict ourselves to Abelian gauge theories. This is enough for the aim of this paper: the determination of the minimal ingredients that allow to build a $\mathcal N=2$ model with two distinct supersymmetry-breaking scales. Note that the electric and magnetic FI terms will play crucial roles in what follows.\footnote{In addition to allowing the breaking of SUSY at two different scales, they also ensure that our theory does not posses a linear anomaly multiplet and thus does not fall into the category of theories covered by the recent no-go theorem in \cite{Antoniadis:2010nj}.}

An Abelian or Maxwell multiplet is described in terms of an $\mathcal N=2$ chiral superfield $\mathcal Z(y,\theta,\tilde\theta)$, with $y^{\mu}=x^{\mu}+i\theta\sigma^{\mu}\bar\theta+i\tilde{\theta}\sigma^{\mu}\bar{\tilde{\theta}}$. The set of Grassmann variables $\tilde{\theta},\bar{\tilde{\theta}}$ is associated with the second supersymmetry. A general chiral superfield can be expanded as follows in terms of $\mathcal N=1$ chiral superfields
\begin{equation}
\label{eqn:genchirexp}
\mathcal Z(y,\theta,\tilde{\theta})=Z(y,\theta)+\sqrt2\tilde{\theta}\omega(y,\theta)+\tilde{\theta}^2\left(\Phi_Z(y,\theta)-\frac{1}{4}\bar{D}^2\bar Z(y,\theta)\right).
\end{equation}
It then describes $16_B+16_F$ off-shell degrees of freedom.

A Maxwell multiplet is expressed in terms of $\mathcal N=1$ superfields as a chiral multiplet $X=\frac{1}{2}\bar D^2V'$, where $V'$ is an arbitrary real superfield, with a vector multiplet $V$ or its supersymmetric field strength $W_{\alpha}$ satisfying the Bianchi identity $DW=\bar D\bar W$. More precisely, the couple $(V',V)$ is the $\mathcal N=2$ supersymmetric extension of the gauge field $A_{\mu}$, while $(X,W_{\alpha})$ is its analog for the curvature $2\partial_{[\mu}A_{\nu]}$.

The Maxwell multiplet contains $8_B+8_F$ fields. Therefore, the general chiral superfield $\mathcal Z$ must be constrained \cite{Lykken:1996xt}:
\begin{equation}
\label{eqn:constrmax}
D\widetilde{D}\mathcal Z+\overline D\overline{\widetilde{D}}\bar{\mathcal Z}=0.
\end{equation}
Note that the usual covariant derivatives $D_{\alpha}$ and $\tilde{D}_{\alpha}$ form a doublet of the $SU(2)_R$ automorphism group of the $\mathcal N=2$ supersymmetry algebra. These equations reduce the chiral superfield $\mathcal Z$ by imposing $\omega_{\alpha}=iW_{\alpha}$, the supersymmetric field strength described above, $Z=X$ and $\Phi_Z=0$, leading to the final form of the Maxwell multiplet\footnote{More rigorously the constraint (\ref{eqn:constrmax}) fixes the superfield $\Phi_Z$ to an arbitrary constant. The important case when this constant does not vanish will be considered later.} $\mathcal W$:
\begin{equation}
\label{eqn:maxsupw}
\mathcal W(y,\theta,\tilde{\theta})=X(y,\theta)+\sqrt2i\tilde{\theta}W(y,\theta)-\frac{1}{4}\tilde{\theta}^2\bar D^2\bar X(y,\theta).
\end{equation}
According to the expression of $\mathcal Z$ (\ref{eqn:genchirexp}) and the previous constraints, the second supersymmetry variations of the Maxwell multiplet $\left(X,V\right)$ are obtained by analogy with $\mathcal N=1$ supersymmetry:
\begin{equation}
\label{eqn:varmax}
\begin{split}
&\tilde{\delta}X=\sqrt2i\tilde{\epsilon}W,\\
&\tilde{\delta}W_{\alpha}=\frac{i}{2\sqrt2}\tilde{\epsilon}_{\alpha}\bar D^2\bar X+\sqrt2\left(\sigma^{\mu}\bar{\tilde{\epsilon}}\right)_{\alpha}\partial_{\mu}X,
\end{split}
\end{equation}
where $\tilde\epsilon$ is an anticommuting infinitesimal parameter. Gauge models are expressed in terms of a single holomorphic prepotential $\mathcal F(\mathcal W)$:
\begin{equation}
\label{eqn:lagprep}
\begin{split}
\mathcal L&=-\frac{i}{2}\int d^2\theta d^2\tilde{\theta}\mathcal F(\mathcal W)  + \textrm{ h.c.}\\
&=\frac{i}{2}\int d^4\theta\left[\bar{\mathcal F}'(\bar X)X-\mathcal F'(X)\bar X\right]-\frac{i}{4}\int d^2\theta\mathcal F''(X)W^2 + \textrm{ h.c.}
\end{split}
\end{equation}
In the second equality, we performed an integration by parts.\footnote{We have dropped total derivatives, and we will systematically ignore such terms in the computations to come.} The ${\mathcal N}=1$ K\"ahler potential~\cite{Wess:1992cp} $K(X,\bar X)$ and the gauge kinetic function $f(X)$ are both defined in terms of the single holomorphic function $\mathcal F$, as can be read from the second line above:
\begin{equation}
\label{eqn:kpgkf}
\begin{array}{cccc}
K(X,\bar X)=\frac{i}{2}\left[\bar{\mathcal F}'(\bar X)X-\mathcal F'(X)\bar X\right] & , & f(X)=-i\mathcal F''(X)& ,
\end{array}
\end{equation}
so that the K\"ahler metric $g_{x\bar x}$ and the metric of the gauge kinetic terms $h_{xx}$ coincide\footnote{$x$ is the (scalar) lowest component of $X$.}: $g_{x\bar x}=h_{xx}=\textrm{Im}\mathcal F''(x)$.

The form of the superpotential allowed by $\mathcal N=2$ supersymmetry is highly restricted. Only electric FI terms can be written \cite{Antoniadis:1995vb,Partouche:1996yp,Marsano:2007mt} in (linearly) realized supersymmetry. Introducing FI parameters breaks the $SU(2)_R$ automorphism. One can choose the direction of this breaking (imagining the $SU(2)_R$ triplet of FI parameters as a three-dimensional vector) to fix some of the parameters to zero \cite{Antoniadis:1995vb}. We will see that the coupling of the Maxwell multiplet to the single-tensor multiplet already selects some preferential direction. Hence we keep a general form of the FI parameters in the discussion here. The FI terms corresponding to the U(1) gauge multiplet $\mathcal W$ can be expressed as:
\begin{equation}
\label{eqn:elfit}
\mathcal L_{FI}=\int d^2\theta d^2\tilde{\theta} \left(\frac{i\xi}{2\sqrt2}\theta\tilde{\theta}-\frac{e}{4}\tilde{\theta}^2\right)\mathcal W + \textrm{ h.c.}=\int d^4\theta\ \xi V- \int d^2\theta\ \frac{e}{4}X + \textrm{ h.c.}\textrm{ ,}
\end{equation}
where $\xi$ is real and $e$ are complex parameters. Notice however \cite{Ambrosetti:2009za} that the relation $X=\frac{1}{2}\bar D^2V'$ implies that the imaginary part of e is multiplied by the curl of a three-index antisymmetric tensor, and disappears from the action.

The most general superpotential compatible with a second linear supersymmetry is then linear in the superfield $X$. A more general possibility arises when one modifies the transformation laws (\ref{eqn:varmax}) by a constant deformation \cite{Antoniadis:1995vb,Partouche:1996yp,Ivanov:1997mt,Marsano:2007mt}. This deformation can be understood from the general expression of the chiral superfield $\mathcal Z$ (\ref{eqn:genchirexp}). Instead of setting $\Phi_z$ to zero, this superfield can be adjusted to a constant value without changing the field content of the Maxwell multiplet: $\Phi_Z=-\frac{iu}{2\kappa}$, with $\kappa$ real and $u$ a phase. The transformation laws (\ref{eqn:varmax}) are therefore deformed as follows:
\begin{equation}
\label{eqn:varmaxdef}
\begin{split}
&\tilde{\delta}X=\sqrt2i\tilde{\epsilon}W,\\
&\tilde{\delta}W_{\alpha}=\frac{i}{2\sqrt2}\tilde{\epsilon}_{\alpha}\bar D^2\bar X+\sqrt2\left(\sigma^{\mu}\bar{\tilde{\epsilon}}\right)_{\alpha}\partial_{\mu}X-\frac{u}{\sqrt2\kappa}\tilde{\epsilon}_{\alpha}.
\end{split}
\end{equation}
The deformation $1/\kappa$ only affects the gaugino $\lambda$ and therefore can be seen as a shift of the auxiliary field $D$ of $W_{\alpha}$. Partial SUSY breaking then originates from the presence of this deformation. The effect of the ``electric" superpotential in (\ref{eqn:elfit}) is to shift the auxiliary field of the magnetic dual theory \cite{Antoniadis:1995vb,Ferrara:1995xi,Partouche:1996yp,Ivanov:1997mt}. The parameter $1/\kappa$ is then the source of a magnetic FI term.

The general form of the gauge model invariant under the deformed variations (\ref{eqn:varmaxdef}) now becomes:
\begin{equation}
\label{eqn:deflagg}
\begin{split}
\mathcal L_{gauge}=&\int d^2\theta d^2\tilde{\theta}\left[-\frac{i}{2}\mathcal F(\mathcal W^a-\frac{iu}{2\kappa_a}\tilde{\theta}^2)+\left(\frac{i\xi_a}{2\sqrt2}\theta\tilde{\theta}-\frac{e_a}{4}\tilde{\theta}^2\right)\mathcal W^a\right]  + \textrm{ h.c.}\\
=&\frac{i}{2}\int d^4\theta\left[\bar{\mathcal F}_{a}(\bar X^b)X^a-\mathcal F_a(X^b)\bar X^a\right]-\frac{i}{4}\int d^2\theta\mathcal F_{ab}(X^c)W^aW^b + \textrm{ h.c.}\\
+&\int d^4\theta\ \xi_aV^a-\int d^2\theta\left(\frac{e_a}{4}X^a+\frac{u}{4\kappa_a}\mathcal F_a(X^b)\right) + \textrm{ h.c.} 
\end{split}
\end{equation}
The gauge index $a$ is introduced to take into account possible additional $U(1)
$  factors\footnote{We adopt the following notation for derivatives with respect to the chiral superfields $X^a$ or its lowest component $x^a$: $\partial_a\partial_b\partial_c\cdots\mathcal F=\mathcal F_{abc\cdots}$}. In addition the form of the metric of the kinetic terms now reads: $h_{ab}=\textrm{Im}\mathcal F_{ab}$.
\subsection{Single-tensor multiplet}
The matter content of our model is first presented in terms of the single-tensor multiplet. This multiplet is composed by a ${\mathcal N}=1$ linear superfield $L$, satisfying $D^2L=\bar D^2L=0$, and a chiral superfield $\Phi$. The field content of a linear multiplet is as follows \cite{Derendinger:1994gx}:
\begin{equation}
\label{eqn:linf}
\begin{split}
L\left(x,\theta,\bar\theta\right)&=C(x)+i\theta\eta(x)-i\bar\theta\bar\eta(x)+\theta\sigma^{\mu}\bar\theta\epsilon_{\mu\nu\rho\sigma}\partial^{\nu}b^{\rho\sigma}(x)\\
&+\frac{1}{2}\theta^2\bar\theta\bar\sigma^{\mu}\partial_{\mu}\eta(x)-\frac{1}{2}\bar\theta^2\theta\sigma^{\mu}\partial_{\mu}\bar\eta(x)+\frac{1}{4}\theta^2\bar\theta^2\Box C(x).
\end{split}
\end{equation}
$C$ is a real scalar, while $b_{\mu\nu}$ is a 2-form field. Notice that this representation of the ($\mathcal N=1$) supersymmetry algebra does not possess any auxiliary field and contains $4_B+4_F$ off-shell degrees of freedom. We will see in the next section that this formulation of supersymmetric matter is well adapted for the description of the Higgs branch. In particular, the two-form plays the role of the would-be Goldstone boson (zero mode) absorbed by the gauge boson of a spontaneously broken symmetry.

Alternatively $L$ can be written in terms of a chiral, spinorial multiplet $\chi_{\alpha}$: $L=D\chi+\bar D\bar\chi$. Hence $L$ is invariant under the supersymmetric gauge transformation \cite{Ambrosetti:2009za}
\begin{equation}
\label{eqn:lingt}
\chi_{\alpha} \longrightarrow \chi_{\alpha}+\frac{i}{4}\bar D^2D_{\alpha}\Delta,
\end{equation}
where $\Delta$ is a real superfield. The main advantage of this formulation of the single-tensor multiplet is that it can be embedded in a $\mathcal N=2$ chiral superfield $\mathcal Y$:
\begin{equation}
\label{eqn:expstm}
\mathcal Y(y,\theta,\tilde\theta)=Y(y,\theta)+\sqrt2\tilde\theta\chi(y,\theta)+\tilde\theta^2\left(-\frac{i}{2}\Phi(y,\theta)-\frac{1}{4}\bar D^2\bar Y(y,\theta)\right).
\end{equation}
The $\mathcal N=2$ supersymmetric version of the gauge transformation (\ref{eqn:lingt}) then becomes $\delta\mathcal Y=-\hat{\mathcal W}$, with $\hat{\mathcal W}$ a Maxwell superfield. According to this, the chiral superfield $\Phi$ is invariant.

The manifestly $\mathcal N=2$ supersymmetric formulation of the single-tensor multiplet requires the introduction of a new chiral multiplet $Y$ that can be set to zero with the help of the gauge transformation. For the $Y$ component, we have $\delta Y=-\frac{1}{2}\bar D^2\Delta'$, with $\Delta'$ a real superfield. However, $Y$ is needed for the $\mathcal N=2$ supersymmetry variations of the single-tensor multiplet $(\chi_{\alpha},\Phi)$ to forming a closed algebra without any additional gauge transformation \cite{Ambrosetti:2009za}. Similarly to the Wess-Zumino gauge one can nevertheless partially fix the gauge in order to stay with the imaginary part of the highest component of $Y$ only, a four-form field $C_{\mu\nu\rho\sigma}$:
\begin{equation}
\label{eqn:redfy}
Y^{gauged}=\frac{i}{24}\theta^2 \epsilon^{\mu\nu\rho\sigma}C_{\mu\nu\rho\sigma}.
\end{equation}

By a computation analogous to (\ref{eqn:varmax}), one finds the following supersymmetry variations in both formulations
\begin{equation}
\label{eqn:ststc}
\begin{array}{ll}
\tilde\delta Y=\sqrt2\tilde\epsilon\chi\textrm{ ,} & \\ 
\tilde\delta \chi_{\alpha}=-\frac{i}{\sqrt2}\tilde\epsilon_{\alpha}\Phi -\frac{1}{2\sqrt2}\tilde\epsilon_{\alpha}\bar D^2\bar Y+\sqrt2i\left(\sigma^{\mu}\bar{\tilde\epsilon}\right)_{\alpha}\partial_{\mu}Y\textrm{ ,} & \tilde\delta L=-\frac{i}{\sqrt2}\left(\tilde\epsilon D\Phi-\bar{\tilde\epsilon}\bar D\bar\Phi\right)\textrm{ ,}\\
\tilde\delta\Phi=2\sqrt2\partial_{\mu}\chi\sigma^{\mu}\bar{\tilde\epsilon}+\frac{i}{\sqrt2}\bar D^2\bar\chi\bar{\tilde\epsilon}\textrm{ ,} & \tilde \delta\Phi=-\sqrt2i\bar{\tilde\epsilon}\bar DL\textrm{ .}
\end{array}
\end{equation}
As the formulation $(L,\Phi)$ is gauge invariant, their transformation laws are independent of the chiral superfield $Y$.

There exists a very simple condition for the invariance of a kinetic action involving the single-tensor multiplet. The variations of $L$ and $\Phi$ in (\ref{eqn:ststc}) lead to the following result. An arbitrary real function $H(L,\Phi,\bar\Phi)$ characterizes a $\mathcal N=2$ supersymmetric model
\begin{equation}
\label{eqn:actst}
\mathcal L_{ST}=\int d^4\theta H\left(L,\Phi,\bar\Phi\right),
\end{equation}
if and only if it fulfills the Laplace equation \cite{Lindstrom:1983rt}: $H_{LL}+2H_{\Phi\bar\Phi}=0$.
\subsection{Chern-Simons interaction and scalar tensor duality}
So far we described the kinetic and self-interaction terms of the gauge and matter sectors of our model. We still need to present their interaction. The Maxwell and the single-tensor multiplets contain both a 2-form field, respectively given by $F_{\mu\nu}$ and $b_{\mu\nu}$. One is thus left with the possibility to build a Chern-Simons interaction $b\wedge F$. Its $\mathcal N=2$ supersymmetric version is expressed in terms of the superfields $\mathcal W$ and $\mathcal Y$ as
\begin{equation}
\label{eqn:csint}
\begin{split}
\mathcal L_{CS}&=-2ig_a\int d^2\theta d^2\tilde{\theta}\mathcal Y\left(\mathcal W^a-\frac{iu}{2\kappa_a}\tilde{\theta^2}\right) + \textrm{ h.c.}\\
&=-g_a\int d^2\theta \left(\Phi X^a+2\chi W^a+\frac{u}{\kappa_a}Y\right) + \textrm{ h.c.}\\
&=2g_a\int d^4\theta LV^a-g_a\int d^2\theta \left(\Phi X^a+\frac{u}{\kappa_a}Y\right) + \textrm{ h.c.}
\end{split}
\end{equation}
The parameters $g_a$ are dimensionless gauge couplings associated to $U(1)$ factors. Similarly to (\ref{eqn:deflagg}), the deformation must be taken into account to make this term invariant under the supersymmetry variations (\ref{eqn:varmaxdef}). In order that (\ref{eqn:csint}) characterize the interaction between a single-tensor and a Maxwell multiplets, invariance under the gauge transformations $\delta\mathcal Y=-\hat{\mathcal W}$, $\delta\mathcal W=0$ is also required. This leads to the condition $u=\pm i$. The sign ambiguity can be absorbed in the definition of $\kappa_a$. So we set $u=i$.

It is crucial to notice that the deformation leaves a dependence on the four-form field of $Y$. Even in the formulation involving the linear multiplet $L$, it cannot be cancelled. It will play an important role in a dual formulation of the model where the single-tensor multiplet is replaced by the hypermultiplet, as we shall discuss now.

The complete Lagrangian of the model in consideration\footnote{As we will see, a superpotential term linear in $\Phi$ is also allowed by (\ref{eqn:ststc}).} is made of (\ref{eqn:deflagg}), (\ref{eqn:actst}) and (\ref{eqn:csint}):
\begin{equation} 
\label{eqn:lagtotst}
\mathcal{L}=\mathcal L_{gauge}+\mathcal L_{ST}+\mathcal L_{CS}.
\end{equation}
Due to the possibility of organizing all the fields in $\mathcal N=2$ chiral superfields, it is manifestly invariant under the ${\mathcal N}=2$ SUSY transformations (\ref{eqn:varmaxdef}) and (\ref{eqn:ststc}). The rewriting of this action in terms of hypermultiplets turns out to be always possible \cite{Lindstrom:1983rt}. However, the inverse duality transformation is only allowed when the hyper-K\"ahler manifold characterizing the target space of the matter sector \cite{AlvarezGaume:1981hm,Bagger:1983tt} possesses a $U(1)$ (shift) isometry.

The duality consists in replacing the linear multiplet by a chiral superfield $\Phi'$ and its complex conjugate $\bar\Phi'$. The matter sector is then described in terms of a K\"ahler potential $K(\Phi'+\bar\Phi',\Phi,\bar\Phi)$ resulting from the Legendre transformation
\begin{equation}
\label{eqn:kahlleg}
K\left(\Phi'+\bar\Phi',\Phi,\bar\Phi\right)=H\left(G,\Phi,\bar\Phi\right)-G\left[\Phi'+\bar\Phi'\right],
\end{equation}
where $H$ is the function defining the kinetic action (\ref{eqn:actst}) and $G$ is the solution of the equation
\begin{equation}
\label{eqn:equleg}
\frac{\partial H\left(G,\Phi,\bar\Phi\right)}{\partial G}=\Phi'+\bar\Phi'-2g_aV^a.
\end{equation}
The $V^a$-dependence comes from the Chern-Simons interaction. By performing the change of variable $\Psi=\exp{\Phi'}$, one recovers the usual form $\bar\Psi e^{-2g_aV^a}\Psi$. Therefore the dualization of the single-tensor multiplet indeed leads to a K\"ahler manifold\footnote{It can be shown that the manifold is actually Ricci-flat. This is a necessary and sufficient condition for a K\"ahler manifold to be hyper-K\"ahler.} with a $U(1)$ isometry generated by a holomorphic Killing vector.
The action is now formulated in a more familiar way. However, the effect of the Legendre transformation is to impose the use of the equations of motion in order to verify $\mathcal N=2$ invariance.
\subsection{Particular model with canonical hyper-K\"ahler manifold and $U(1)\times U(1)$ gauge group}
So far we only specified that our model contains one $\mathcal N=2$ matter multiplet expressed as a single-tensor or a hyper- multiplet, and $n$ $U(1)$ Maxwell superfields. From now on, we restrict it by imposing that the K\"ahler potential be canonical:
\begin{equation}
\label{eqn:kahlpotcan}
K\left(Q^u,\bar Q^{\bar u}\right)=\bar Q^1Q^1+\bar Q^2Q^2.
\end{equation}
We changed the notation of the two chiral multiplets of the hypermultiplet\footnote{The field expansion of $Q^u$ is written as $Q^u=q^u+\sqrt2\theta\chi^u+\theta^2F^u$.}: $(Q^1,Q^2)$. These two superfields transform under conjugate representations of the gauge group. $\Phi$ will instead be used only in the single-tensor formalism. The model dual to (\ref{eqn:lagtotst}) with the particular K\"ahler potential (\ref{eqn:kahlpotcan}) is thus given by
\begin{equation}
\label{eqn:lagtothyp}
\begin{split}
\mathcal{L}&=\int d^4\theta \left[\bar{Q}^1 e^{-2g_aV^a}Q^1 + \bar{Q}^2 e^{2g_aV^a}Q^2+\frac{i}{2}\left(\bar{\mathcal{F}}_{a}X^a-\mathcal{F}_a\bar X^a\right)
+\xi_aV^a\right]\\
&+\int d^2\theta\left[-\frac{i}{4}\mathcal{F}_{ab}W^{a}W^{b}+\left(m+\sqrt2i g_aX^a\right)Q^1Q^2-\frac{e_a}{4}X^a-\frac{i}{4\kappa_a}\mathcal F_a\right.\\
&\left.\hspace{1.7cm}-\frac{ig_a}{\kappa_a}Y\right] + \textrm{ h.c.},
\end{split}
\end{equation}
where $m$ is a mass parameter. In the absence of gauge interactions, the mass term is the only superpotential compatible with (\ref{eqn:kahlpotcan}) and $\mathcal N=2$ supersymmetry \cite{Mezincescu:1983rn}. Ref. \cite{Bagger:2006hm} provides an interesting geometrical interpretation of this result. The invariance of the action under $\mathcal N=2$ supersymmetry is checked by proceeding to the inverse duality transformation in order to obtain the function $H(L,\Phi,\bar\Phi)$. The dualization is done once an appropriate change of variable is performed \cite{Ambrosetti:2009za}:
\begin{equation}
\label{eqn:changvar}
Q^1=\sqrt{\frac{\Phi}{\sqrt2}}e^{-\Phi'} \textrm{ , } Q^2=i\sqrt{\frac{\Phi}{\sqrt2}}e^{\Phi'}.
\end{equation}
Then a Legendre transformation leads to the expression
\begin{equation}
\label{eqn:parth}
H\left(L,\Phi,\bar\Phi\right)=\sqrt{L^2+2\Phi\bar\Phi}-L\ln{\left(L+\sqrt{L^2+2\Phi\bar\Phi}\right)}.
\end{equation}
It is now straightforward to verify that this function satisfies the Laplace equation, and this completes the proof of the (linear) supersymmetry invariance, since the superpotential term $\frac{im}{\sqrt2}\Phi$ is invariant according to the transformation laws (\ref{eqn:ststc}).

We want to show that the minimal value of $n$ in order that (\ref{eqn:lagtothyp}) be invariant under the deformed supersymmetry is $n=2$. Indeed several works \cite{Partouche:1996yp,Ivanov:1997mt} pointed out difficulties in building a theory involving a hypermultiplet charged under a single $U(1)$ and for which one supersymmetry is nonlinearly realized (spontaneously broken). This  is understood as follows in our approach: the difficulties come from the Y-dependent term in the hypermultiplet formulation. The Lagrangian (\ref{eqn:lagtothyp}) depends linearly on the 4-form field and its equation of motion implies the constraint:
\begin{equation}
\label{eqn:addconstr}
\frac{g_a}{\kappa_a}=0.
\end{equation}
The dependence of the non-dynamical fields is therefore completely cancelled.\footnote{This phenomenon is known from D-branes dynamics, where an analog constraint ensures tadpole cancellation and vanishing of the 3-brane charge associated to the 4-form gauge potential Y.}

In the presence of a single $U(1)$, either the hypermultiplet is not charged $(g=0)$ and we are then back\footnote{An equivalent statement was already done in \cite{Fujiwara:2005kf} where it is shown that there is no difficulty preventing the invariance under a nonlinear supersymmetry of a model with hypermultiplets in the adjoint representation of the gauge group.} to the model of \cite{Antoniadis:1995vb} or the second supersymmetry is linearly realized $(\frac{1}{\kappa}=0)$. Consistency of the equations of motion thus implies that the hypermultiplet be at least charged under two $U(1)$ factors, and that magnetic FI terms be associated with these Abelian gauge groups. This result was first derived in \cite{Itoyama:2007kk} in the framework of harmonic superspace. We will then focus in this note on the minimal consistent case, that is a $U(1)\times U(1)$ gauge group.

Note that the $\mathcal N=2$ supersymmetry invariance of the Lagrangian (\ref{eqn:lagtothyp}) does not constrain the form of the electric and magnetic FI coefficients in the hypermultiplet formalism. In particular, $\kappa_a$ could be taken complex. However, as mentioned earlier, one can use the $SU(2)_R$ automorphism group, to eliminate some of their components. We want to preserve the restriction imposed by the single-tensor formalism and take $\kappa_a$ real while keeping $e_a$ complex. The duality transformation so performed corresponds to a choice of frame in the space of FI coefficients different from the one in the APT model \cite{Antoniadis:1995vb} but equivalent.
\section{Vacua of the model\label{vac}}
We are now able to analyze the vacua of the model. We will proceed slightly differently for the Coulomb phase, where the $U(1)\times U(1)$ gauge group is preserved, and for the Higgs phase where the gauge group is broken. The Coulomb phase can only be studied in the hypermultiplet formalism, since it is defined by the vanishing of the vacuum expectation value (VEV) of the matter scalar fields: $\langle q^u\rangle=0$. The field redefinition (\ref{eqn:changvar}) is indeed ill-defined at the origin. The single-tensor formalism can then describe only the Higgs branch. In fact, in this phase, the analysis turns out to be easier in the single-tensor formulation.
\subsection{Coulomb phase}
The relevant part of the Lagrangian (\ref{eqn:lagtothyp}) for the study of the vacuum structure is the scalar potential \cite{Wess:1992cp}:
\begin{equation}
\label{eqn:scalpothyp}
\begin{split}
V_S&=\left|m_{eff}\right|^2\left(\left|q^1\right|^2+\left|q^2\right|^2\right)+h^{ab}\left(-\sqrt2i g_a q^1q^2+\frac{e_a}{4}+\frac{i\mathcal F_{ac}}{4\kappa_c}\right)\left(\sqrt2i g_b \bar q^{\bar1}\bar q^{\bar2}+\frac{\bar e_b}{4}-\frac{i\bar{\mathcal F}_{bd}}{4\kappa_d}\right)\\
&+\frac{1}{8}h^{ab}\left[-2g_a\left(\left|q^1\right|^2-\left|q^2\right|^2\right)+\xi_a\right]\left[-2g_b\left(\left|q^1\right|^2-\left|q^2\right|^2\right)+\xi_b\right],
\end{split}
\end{equation}
where $m_{eff}$ is a shortcut for the following expression,
\begin{equation}
\label{eqn:meff}
m_{eff}=m+\sqrt2ig_ax^a\, ,
\end{equation}
while $h^{ab}$ is the inverse metric of the special-K\"ahler manifold \cite{deWit:1984pk,Cremmer:1984hj} of the gauge sector. The possible vacua of the theory correspond to points of the target space that are solutions of the stationarity conditions (minima of the scalar potential). These are given by the first derivative of $V_S$ with respect to the scalar fields of the model $q^u$, $u=1,2$ and $x^a$, $a=1,2$:
\begin{equation}
\label{eqn:statcondhyp}
\begin{split}
&\left|m_{eff}\right|^2\bar q^{\bar1}-\sqrt2i g_aF^{x^a}q^2+g_aD^a\bar q^{\bar1}=0\, ,\\
&\left|m_{eff}\right|^2\bar q^{\bar2}-\sqrt2i g_aF^{x^a}q^1-g_aD^a\bar q^{\bar2}=0\, ,\\
&\frac{i}{2}\mathcal{F}_{abc}\left[F^{x^b}\left(\bar F^{\bar x^c}+\frac{1}{2\kappa_c}\right)+\frac{1}{2}D^bD^c\right]-\sqrt2i g_a\left(F^{q^1}q^2+F^{q^2}q^1\right)=0\, .
\end{split}
\end{equation}
The auxiliary fields of the hyper- and the Maxwell multiplets, respectively $F^{q^u}$, $F^{x^a}$ and $D^a$, are expressed for vanishing fermions and gauge fields as
\begin{equation}
\label{eqn:auxfieldshyp}
\begin{split}
&F^{q^1}=-\bar{m}_{eff}\bar q^{\bar2} \textrm{ , } F^{q^2}=-\bar{m}_{eff}\bar q^{\bar1},\\
&F^{x^a}=h^{ab}\left(\sqrt2ig_b\bar q^{\bar1}\bar q^{\bar2}+\frac{\bar e_b}{4}-\frac{i\textrm{Re}\mathcal{F}_{bc}}{4\kappa_c}\right)-\frac{1}{4\kappa_a},\\
&D^a=-\frac{1}{2}h^{ab}\left[-2g_b\left(\left|q^1\right|^2-\left|q^2\right|^2\right)+\xi_b\right].
\end{split}
\end{equation}
According to (\ref{eqn:auxfieldshyp}), the resolution of the stationarity conditions is simplified in the Coulomb phase due to the vanishing of the matter scalar VEVs. The first two equations of (\ref{eqn:statcondhyp}) are trivially satisfied, while the third one can be simplified to
\begin{equation}
\label{eqn:statcoul}
\mathcal{F}_{abc}\left[F^{x^b}\left(\bar F^{\bar x^c}+\frac{1}{2\kappa_c}\right)+\frac{1}{2}D^bD^c\right]=0.
\end{equation}
To proceed further with the analysis of the vacua, the prepotential $\mathcal F\left(X^1,X^2\right)$ must be specified. There is actually one particular case, where its precise form is not required. It corresponds to the vanishing of the expression in brackets above. We will show in section \ref{gold} that in such a vacuum supersymmetry is partially broken.\footnote{Note that in the Coulomb phase $\langle D\rangle\ne 0$ for non-vanishing $\xi$ and thus ${\mathcal N}=2$ SUSY is always broken.}

\subsection{Higgs phase}
The hypermultiplet formulation (\ref{eqn:scalpothyp}), (\ref{eqn:statcondhyp}) and (\ref{eqn:auxfieldshyp}) is perfectly valid for the analysis of the Higgs phase, as well (on-shell). However, the resolution of the stationarity conditions is more illuminating in the dual theory. We recall the dual form of the Lagrangian (\ref{eqn:lagtothyp}):
\begin{equation}
\label{eqn:lagtotstdev}
\begin{split}
\mathcal{L}&=\int d^4\theta \hspace{0.1cm}\hspace{-0.1cm}\left[\sqrt{L^2+2\Phi\bar\Phi}-L\ln\left(L+\sqrt{L^2 + 2\Phi\bar\Phi}\right)+2g_aLV^a\right]+\mathcal L_{gauge}\\
&+\int d^2\theta\frac{i}{\sqrt2}\left(m+\sqrt2i g_aX^a\right)\Phi + \textrm{ h.c.}
\end{split}
\end{equation}
Its field expansion shows that, while being equivalent to the model with a flat hyper-K\"ahler manifold, the kinetic terms of the matter fields acquire a non-trivial metric\footnote{According to (\ref{eqn:lagtotstdev}), $L$ and $\Phi$ are of dimension 2, so are their lowest scalar components, $C$ and $\phi$. To have a dimensionless metric $g_{CC}=g_{\phi\bar\phi}$, one could introduce an arbitrary mass scale in the definition of the function (\ref{eqn:parth}). However one can check that this scale disappears in all physically relevant quantities. So for the sake of simplicity, we chose to not introduce it.}:
\begin{equation}
\label{eqn:stmet}
g_{CC}=g_{\phi\bar\phi}=\frac{1}{2\sqrt{C^2+2\left|\phi\right|^2}}.
\end{equation}

The inverse metric appears in the scalar potential (\ref{eqn:scalpothyp}), which now reads
\begin{equation}
\label{eqn:scalpotst}
\begin{split}
V_S&=\sqrt{C^2+2\left|\phi\right|^2}\left|m_{eff}\right|^2+h^{ab}\left(g_a\phi+\frac{e_a}{4}+\frac{i\mathcal F_{ac}}{4\kappa_c}\right)\left(g_b\bar\phi+\frac{\bar e_b}{4}-\frac{i\bar{\mathcal F}_{bd}}{4\kappa_d}\right)\\
&+\frac{1}{8}h^{ab}\left(2g_aC+\xi_a\right)\left(2g_bC+\xi_b\right).
\end{split}
\end{equation}
The form of $V_S$ could be deduced without the full field expression (\ref{eqn:lagtotstdev}). The change of variables (\ref{eqn:changvar}) indeed implies the following relations between the different scalar fields: $C=-\left|q^1\right|^2+\left|q^2\right|^2$ and $\phi=-\sqrt2iq^1q^2$. Starting from (\ref{eqn:scalpothyp}) and imposing invariance of the scalar potential under this field redefinition, one easily recovers (\ref{eqn:scalpotst}).

We now proceed along the same line of reasoning as for the Coulomb phase. The stationarity conditions are given by
\begin{equation}
\label{eqn:statcondst}
\begin{split}
&\frac{C}{2\left(C^2+2\left|\phi\right|^2\right)^{3/2}}\left |F^{\phi}\right|^2-g_aD^a=0\, ,\\
&\frac{\bar\phi}{2\left(C^2+2\left|\phi\right|^2\right)^{3/2}}\left|F^{\phi}\right|^2+g_aF^{x^a}=0\, ,\\
&\frac{i}{2}\mathcal{F}_{abc}\left[F^{x^b}\left(\bar F^{\bar x^c}+\frac{1}{2\kappa_c}\right)+\frac{1}{2}D^bD^c\right]+g_aF^{\phi}=0\, .
\end{split}
\end{equation}
The first two equations correspond, respectively, to the derivatives of $V_S$ with respect to $C$ and $\phi$. The scalar field expressions of the auxiliary fields now become
\begin{equation}
\label{eqn:auxfieldsst}
\begin{split}
&F^{\phi}=i\sqrt{2C^2+4\left|\phi\right|^2}\bar{m}_{eff},\\
&F^{x^a}=h^{ab}\left(g_b\bar\phi+\frac{\bar e_b}{4}-\frac{i\textrm{Re}\mathcal{F}_{bc}}{4\kappa_c}\right)-\frac{1}{4\kappa_a},\\
&D^a=-\frac{1}{2}h^{ab}\left(2g_b C+\xi_b\right).
\end{split}
\end{equation}
As already noticed, there is one auxiliary field less in comparison with the hypermultiplet formulation. This is the first simplification of the single-tensor multiplet formalism.

By inspection of the field redefinition described above, one finds that in the Higgs phase, $\langle C\rangle$ and $\langle\phi\rangle$ cannot simultaneously vanish. Considering a generic choice of electric FI coefficients in (\ref{eqn:deflagg}), where both types are different from zero, one can simplify the system (\ref{eqn:statcondst}) by taking linear combinations of the equations. Multiplying the first line by $\bar\phi$, the second by $C$ and then taking the difference, $\langle\phi\rangle$ can be written in terms of $\langle C\rangle$ and $\langle x^a\rangle$. Putting the new relation together with the first and the third of (\ref{eqn:statcondst}), the stationarity conditions are rewritten as
\begin{equation}
\label{eqn:phiexpcx}
\begin{split}
&C=-\frac{1}{2g_ch^{cd}g_d}\left(g_ah^{ab}\xi_b\pm\frac{2\left|m_{eff}\right|^2}{\sqrt{1+2\left|\alpha\right|^2}}\right) \textrm{ , } \alpha=\frac{g_ah^{ab}\left(e_b+\frac{i\textrm{Re}\mathcal F_{bc}}{\kappa_c}\right)}{2g_ch^{cd}\xi_d}\textrm{ ,}\\
&\phi=\alpha C\textrm{ ,}\\
&\frac{i}{2}\mathcal F_{abc}\left[F^{x^b}\left(\bar F^{\bar x^c}+\frac{1}{2\kappa_c}\right)+\frac{1}{2}D^bD^c\right]+g_aF^{\phi}=0\, ,
\end{split}
\end{equation}
where we assumed that $g_ah^{ab}\xi_b$ does not vanish. This is generically true. The two signs in the first line above correspond respectively to positive and negative values of $\langle C\rangle$.

The last equation resembles (\ref{eqn:statcoul}) and thus same difficulties may appear in their resolution. The VEVs of the matter scalars $C$ and $\phi$ only depend on $\langle x^a\rangle$. Inserting them in the third condition leads to a rather intricate system of equations. The only trivial solution is related to partial supersymmetry breaking as already pointed out in our discussion of the Coulomb phase.

There is a second advantage to work with the single-tensor formulation. The identification of the \lq\lq would-be" Goldstone bosons absorbed by the gauge fields of spontaneously broken symmetries is made easier. In the hypermultiplet formalism, these are linear combinations of all scalars with coefficients determined by the Killing vectors of the K\"ahler manifold generating the broken symmetries. In the single-tensor formalism on the other hand, it turns out that this role is played by the 2-form field $b_{\mu\nu}$. To see this, note that the relevant part of the Lagrangian (\ref{eqn:lagtotstdev}) contains the terms involving the $A_{\mu}$ and $b_{\mu\nu}$ kinetic terms and their interaction:
\begin{equation}
\label{eqn:partlagwg}
\begin{split}
\mathcal{L}\supset &-\frac{h_{11}}{4}F_{\mu\nu}^1F^{1\mu\nu}-\frac{h_{12}}{2}F_{\mu\nu}^1F^{2\mu\nu}-\frac{h_{22}}{4}F_{\mu\nu}^2F^{2\mu\nu}+\frac{1}{4\sqrt{C^2+2\left|\phi\right|^2}}\varepsilon_{\mu\nu\rho\sigma}\partial^{\nu}b^{\rho\sigma}\varepsilon^{\mu\alpha\beta\gamma}\partial_{\alpha}b_{\beta\gamma}\\
&+\varepsilon_{\mu\nu\rho\sigma}\partial^{\nu}b^{\rho\sigma}\left(g_1A^{1\mu}+g_2A^{2\mu}\right)\\
=&-\frac{1}{4}F_{\mu\nu}^{*1}F^{*1\mu\nu}-\frac{1}{4}F_{\mu\nu}^{*2}F^{*2\mu\nu}+\frac{1}{4\sqrt{C^2+2\left|\phi\right|^2}}\varepsilon_{\mu\nu\rho\sigma}\partial^{\nu}b^{\rho\sigma}\varepsilon^{\mu\alpha\beta\gamma}\partial_{\alpha}b_{\beta\gamma}\\
&+\varepsilon_{\mu\nu\rho\sigma}\partial^{\nu}b^{\rho\sigma}\left(g_1^*A^{*1\mu}+g_2^*A^{*2\mu}\right).
\end{split}
\end{equation}
In the right hand side (RHS) above, we redefined gauge fields in order to obtain canonically normalized kinetic terms. An additional field redefinition is required to show how a mass term for the gauge fields appears:
\begin{equation}
\label{eqn:gaugbred}
A_{\mu}^+\equiv \frac{1}{\sqrt{g_1^{*2}+g_2^{*2}}}\left(g_1^*A_{\mu}^{*1}+g_2^{*}A_{\mu}^{*2}\right) \textrm{ , } A_{\mu}^-\equiv \frac{1}{\sqrt{g_1^{*2}+g_2^{*2}}}\left(g_2^*A_{\mu}^{*1}-g_1^{*}A_{\mu}^{*2}\right),
\end{equation}
leading to
\begin{equation}
\label{eqn:redeflagf}
\begin{split}
\mathcal{L}\supset&-\frac{1}{4}F_{\mu\nu}^{+}F^{+\mu\nu}-\frac{1}{4}F_{\mu\nu}^{-}F^{-\mu\nu}-\left(g_1^{*2}+g_2^{*2}\right)\sqrt{C^2+2\left|\phi\right|^2}A_{\mu}^+A^{+\mu}\\
&+\frac{1}{4\sqrt{C^2+2\left|\phi\right|^2}}\left[\varepsilon_{\mu\nu\rho\sigma}\partial^{\nu}b^{\rho\sigma}+2\sqrt{\left(g_1^{*2}+g_2^{*2}\right)\left(C^2+2\left|\phi\right|^2\right)}A_{\mu}^+\right]^2.
\end{split}
\end{equation}
The term in brackets can be set to zero by choosing the unitary gauge. The 2-form is indeed eaten by the gauge field $A^+_{\mu}$ which acquires a mass $m^2_{A^+_{\mu}}=2\left(g^{*2}_1+g^{*2}_2\right)\sqrt{\langle C\rangle^2+2\left|\langle\phi\rangle\right|^2}$. 

This becomes more manifest in the dual representation where the antisymmetric tensor is replaced by a (pseudo)scalar. More precisely, one replaces in (\ref{eqn:redeflagf}) $\varepsilon_{\mu\nu\rho\sigma}\partial^{\nu}b^{\rho\sigma}$ by a general vector field $H_{\mu}$ and a Lagrange multiplier $\varphi$, interchanging equations of motion with Bianchi identities:
\begin{equation}
\label{eqn:2}
\begin{split}
\mathcal{L}\supset-
\frac{1}{4}F^+_{\mu\nu}F^{+\mu\nu}-\frac{1}{4}F^-_{\mu\nu}F^{-\mu\nu}+\frac{1}{4\sqrt{C^2+2\left|\phi\right|^2}}H_{\mu}H^{\mu}
-H^{\mu}\left(\partial_{\mu}\varphi-\sqrt{g^{*2}_1+g^{*2}_2}A^+_{\mu}\right).
\end{split}
\end{equation}
Integrating now over $H_{\mu}$ (instead over $\varphi$), one obtains a more familiar form:
\begin{equation}
\label{eqn:3}
\mathcal{L}=-\frac{1}{4}F^+_{\mu\nu}F^{+\mu\nu}-\frac{1}{4}F^-_{\mu\nu}F^{-\mu\nu}-\sqrt{C^2+2\left|\phi\right|^2}\left(\sqrt{g^{*2}_1+g^{*2}_2}A^+_{\mu}-\partial_{\mu}\varphi\right)^2,
\end{equation}
where the unitary gauge corresponds to choosing $\varphi=0$.

One is thus left with a single massless gauge field $A^-_{\mu}$, indicating that the gauge group $U(1)\times U(1)$ is broken to $U(1)$. In order to break the full gauge group, one has to introduce a second single-tensor multiplet, or equivalently a hypermultiplet with different $U(1)$ charges than the first one.
\section{Goldstini and SUSY-breaking scales\label{gold}}
Partial breaking of global supersymmetry may seem impossible at the level of the supercharge algebra. A quick analysis shows that as soon as one supersymmetry is broken, all are broken too \cite{Witten:1981nf}. A loophole to this argument \cite{Lopuszanski:1978df,Antoniadis:1995vb,Ferrara:1995xi,Partouche:1996yp} is based on the presence of magnetic FI terms. When the second supersymmetry is deformed as we discussed in section \ref{mod}, the supercurrent algebra develops a constant term proportional to the deformation $1/\kappa$. The supercharges are therefore ill-defined and the previous argument cannot be applied. It is then not surprising that the existence of two different supersymmetry-breaking scales is related to the inclusion of a magnetic FI term in the action, as we will show below.
\subsection{Goldstini}
An efficient way for determining whether partial or full SUSY breaking occurs is to compute the two goldstini of the model, the fermionic Goldstone particles arising when supersymmetry is spontaneously broken, and check whether these two combinations of fermions are linearly independent. Should this occur, $\mathcal N=2$ supersymmetry is broken to $\mathcal N=0$. Otherwise, SUSY is only partially broken.

The expression of the two goldstini $\eta_1$, $\eta_2$ is derived from the scalar dependence of the fermionic supersymmetry variations. These are given in appendix \ref{app:varhyp}. The ones relevant for the current purpose become on the vacuum: $\delta\chi^u=\sqrt2\epsilon F^{q^u}$, $\delta\psi^{x^a}=\sqrt2\epsilon F^{x^a}$ and $\delta\lambda^a=i\epsilon D^a$ for the first supersymmetry, $\tilde{\delta}\chi^{u}=\sqrt{2}\tilde{\epsilon}\tilde{F}^{q^u}$, $\tilde\delta\psi^{x^a}=i\tilde\epsilon D^a$ and $\tilde\delta\lambda^a=\sqrt2\tilde\epsilon\bar F^{\bar x^a}+\frac{1}{\sqrt2\kappa_a}\tilde\epsilon$ for the second (deformed) supersymmetry. The fields $\tilde F^{q^u}$ are introduced in appendix \ref{app:varhyp}. We repeat their form here for self-completeness:
\begin{equation}
\label{eqn:tildef}
\tilde F^{q^1}=-\bar{m}_{eff} q^{1} \textrm{ , } \tilde F^{q^2}=\bar{m}_{eff} q^{2}.
\end{equation}
As pointed out in section \ref{mod}, the deformation only acts on the second supersymmetry transformation of the gaugini $\lambda^a$.

The two goldstini of our model are finally found to be given by
\begin{equation}
\label{eqn:goldstini}
\begin{split}
&\eta_1=\frac{1}{N_1}\left(g_{u\bar v}\bar F^{\bar q^{\bar v}}\chi^u+h_{ab}\bar F^{\bar x^a}\psi^{x^b}-\frac{i}{\sqrt2}h_{ab}D^a\lambda^b\right),\\
&\eta_2=\frac{1}{N_2}\left(g_{u\bar v}\bar{\tilde F}^{\bar q^{\bar v}}\chi^u-\frac{i}{\sqrt2}h_{ab}D^a\psi^{x^b}+h_{ab}\left(F^{x^a}+\frac{1}{2\kappa_a}\right)\lambda^b\right).
\end{split}
\end{equation}
The hyper-K\"ahler metric $g_{u\bar v}$ is simply the trivial metric $\delta_{u\bar v}$ here. $N_1$ and $N_2$ are dimension two normalization constants such that the dimension of the goldstini be $3/2$ like usual spin $1/2$ particles:
\begin{equation}
\label{eqn:normconst}
\begin{split}
&N_1=\sqrt{g_{u\bar v}F^{q^u}\bar F^{\bar q^{\bar v}}+h_{ab}\left[F^{x^a}\bar F^{\bar x^b}+\frac{1}{2}D^aD^b\right]}\, ,\\
&N_2=\sqrt{g_{u\bar v}\tilde F^{q^u}\bar{\tilde F}^{\bar q^{\bar v}}+h_{ab}\left[\left(F^{x^a}+\frac{1}{2\kappa_a}\right)\left(\bar F^{\bar x^b}+\frac{1}{2\kappa_b}\right)+\frac{1}{2}D^aD^b\right]}=\sqrt{N_1^2+E}\, .
\end{split}
\end{equation}
Comparing the expressions of $F^{q^u}$ (\ref{eqn:auxfieldshyp}) and $\tilde F^{q^u}$ (\ref{eqn:tildef}), one immediately observes that in our model $\delta_{u\bar v}F^{q^u}\bar F^{\bar q^{\bar v}}=\delta_{u\bar v}\tilde F^{q^u}\bar{\tilde F}^{\bar q^{\bar v}}$. The dependence on the hypermultiplets is then the same for both normalization constants. By inspection of the previous formulas, $E$ is given by
\begin{equation}
\label{eqn:eleccontr}
E=\frac{e_a+\bar e_a}{8\kappa_a}.
\end{equation}
This notation will turn out to be convenient for the expression of the two SUSY breaking scales below.

A verification of the form of the goldstini is to compute their mass. Using the fermion mass matrix given in appendix (\ref{eqn:fermmass}) and the stationarity conditions, one can easily see that they are massless. Notice however that the computation of the second Goldstino mass requires the constraint (\ref{eqn:addconstr}). In a single $U(1)$ model with charged matter, this constraint cannot be satisfied as already discussed. In that case, while global $\mathcal N=2$ supersymmetry can be made manifest, the second Goldstino is massive. The origin of this super-Higgs mechanism in global supersymmetry and the connection to the non-dynamical four-form field $C_{\mu\nu\rho\sigma}$ was pointed out in \cite{Ambrosetti:2009za}.
\subsection{SUSY-breaking scales\label{susybs}}
The order parameter of $\mathcal N=1$ supersymmetry breaking is given by the VEV of the scalar potential, which can be interpreted as the square norm of the vector of auxiliary fields: $V_S=g_{i\bar\jmath}F^{i}\bar F^{\bar\jmath}+\frac{1}{2}h_{ab}D^aD^b$. Another quantity that may play the role of order parameter is the variation of the Goldstino $\delta\eta_1$. This must actually be equivalent to $\langle V_S\rangle$. The expression (\ref{eqn:goldstini}) can thus be alternatively verified by demanding that $\delta\eta_1$ depend only on the combination of auxiliary fields $V_S$. According to the transformation laws (\ref{eqn:compvarn1}), one finds:
\begin{equation}
\label{eqn:vargold1}
\delta\eta_1=\sqrt2N_1=\sqrt{2\langle V_S\rangle}.
\end{equation}

The VEV of the scalar potential also characterizes the scale at which global supersymmetry is broken: $\Lambda_{SUSY}\sim \langle V_S\rangle^{1/4}$. Computing it from the variation of the Goldstino is well adapted for the extension to $\mathcal N=2$ SUSY. One may expect two scales in this case, one for each supersymmetry. To see this, the variation $\delta\eta_1$ is generalized to a $2\times2$ matrix\footnote{According to the notation of section \ref{mod}, $\delta_1=\delta$ and $\delta_2=\tilde{\delta}$.} $\delta_i\eta_j$, $i,j=1,2$. Its eigenvalues therefore define the two supersymmetry-breaking scales:
\begin{equation}
\label{eqn:susyscales}
\left(\Lambda^{1,2}_{SUSY}\right)^2=\frac{N_1+\sqrt{N_1^2+E}}{\sqrt2}\pm\frac{1}{2}\sqrt{2\left(N_1-\sqrt{N_1^2+E}\right)^2+\frac{\left(\frac{h_{ab}D^b}{\kappa_a}\right)^2}{N_1\sqrt{N_1^2+E}}}.
\end{equation}
One immediately sees that magnetic FI coeffcients $1/\kappa_a$ play a crucial role in the lifting of degeneracy of the two scales\footnote{To express these scales in the single-tensor formalism, one must simply replace the term $g_{u\bar v}F^{q^u}\bar F^{\bar q^{\bar v}}$ by $g_{\phi\bar\phi}F^{\phi}\bar F^{\bar\phi}$.}. In the absence of the deformation they become equal, as the action satisfies again the hypotheses of the no-go theorem that asserts that all supersymmetries are simultaneously broken. A trivial consequence is that the two SUSY-breaking scales are identical in that case\footnote{Notice that the no-go theorem is only valid in global supersymmetry. Partial SUSY breaking in local supersymmetry was first demonstrated in \cite{Ferrara:1995gu}. In supergravity theories SUSY-breaking scales are defined by the mass of the gravitini \cite{Cecotti:1984rk}, which are not necessarily equal.}.

It is more illuminating to consider the two types of contributions $E$ and $\frac{h_{ab}D^b}{\kappa_a}$ separately. According to (\ref{eqn:eleccontr}) the first deals with the $e_a$ part of the electric FI coefficients, while the second takes account of the effect of $\xi_a$. After having imposed a $D$-flatness condition, one finds
\begin{equation}
\label{eqn:dflatlim}
\left(\Lambda^{1}_{SUSY}\right)^2=\sqrt2N_1 \textrm{ , } \left(\Lambda^{2}_{SUSY}\right)^2=\sqrt{2\left(N_1^2+E\right)}.
\end{equation}
The form of $E$ (\ref{eqn:eleccontr}) only involves the real part of $e_a$. Hence a possibility to get two widely separate scales is to impose $E\gg N_1^2$. In case the constants $e_a$ are imaginary, the two SUSY breaking scales become
\begin{equation}
\label{eqn:eimaglim}
\left(\Lambda^{1,2}_{SUSY}\right)^2=\frac{\sqrt2}{N_1}\left[g_{u\bar v}F^{q^u}\bar F^{\bar q^{\bar v}}+h_{ab}\textrm{Im}F^{x^a}\textrm{Im}F^{x^b}+h_{ab}\left(\frac{1}{4\kappa_a}\pm\frac{D^a}{\sqrt2}\right)\left(\frac{1}{4\kappa_b}\pm\frac{D^b}{\sqrt2}\right)\right].
\end{equation}
In order to break partially supersymmetry, it is then not sufficient to introduce only imaginary electric coefficients $e_a$ in addition to the magnetic ones. Interpreting this statement in terms of triplets  of FI terms, one finds that partial SUSY breaking is not possible when the electric and magnetic triplets are parallel\footnote{We recall that the invariance of the Chern-Simons Lagrangian (\ref{eqn:csint}) under gauge transformations automatically selects imaginary magnetic FI coefficients.}. This result was derived in \cite{Ferrara:1995xi}, where it has been shown that the modification of the current algebra is proportional to the cross product of these two types of vectors.

On a vacuum breaking partially supersymmetry one scale vanishes, since the two goldstini are linearly dependent and therefore the matrix $\delta_i\eta_j$ has rank 1. For example, in the case described by (\ref{eqn:eimaglim}) each of the three terms in brackets must vanish due to the positive-definiteness of the metrics $g_{u\bar v}$ and $h_{ab}$. Partial SUSY-breaking vacua are then specified by the VEVs
\begin{equation}
\label{eqn:vevpartb}
\langle F^{x^a}\rangle =-\frac{1}{4\kappa_a} \textrm{ , } \langle D^a\rangle =\mp\frac{1}{2\sqrt2\kappa_a} \textrm{ , } \langle F^{q^u}\rangle=\langle \tilde F^{q^u}\rangle=0.
\end{equation}
They are solution of the stationarity conditions (\ref{eqn:statcondhyp}). The prepotential $\mathcal F(X^1,X^2)$ does not need to be fully specified to solve them because the term in brackets multiplying its third derivatives vanishes.

By comparing these VEVs with (\ref{eqn:goldstini}), we deduce that two possible relations between the goldstini arise: $\eta_2=\mp i\eta_1$. The supersymmetry preserved on the vacuum is generated by one of the two following combinations of supercharges: $\mathcal Q\pm i\tilde{\mathcal Q}$. One can usually discriminate between the two linear combinations by imposing the positive-definiteness of $h_{ab}$, which selects one of the two signs of $\langle D^a\rangle$ in (\ref{eqn:vevpartb}), respectively.
\subsection{Toy model with two SUSY-breaking scales}
As a step towards constructing a realistic $\mathcal N=2$ supersymmetric model, it is useful to allow the two supersymmetries to be broken at parametrically distinct energies. In many phenomenologically viable extensions of the SM \cite{Haber:1993wf}, one has to be broken at $\mathcal O(100)$ GeV, for stabilizing the electroweak scale and solving the hierarchy problem, while the other could be broken at a much higher energy, separated from the electroweak scale by several orders of magnitude.

We would like to investigate the intermediate regime $\Lambda_{int}$ defined as $\Lambda^1_{SUSY}\ll\Lambda_{int}\ll\Lambda^2_{SUSY}$ in a simple toy model. We make a detailed analysis of the Coulomb phase of our model (\ref{eqn:lagtothyp}) with a separable prepotential:
\begin{equation}
\label{eqn:partprep}
\mathcal F(X^1,X^2)=f_1\left(X^1\right)+f_2\left(X^2\right).
\end{equation}
The non-abelian extension of this particular model was considered in \cite{Itoyama:2007kk}, where it was shown that under an appropriate choice of the relative sign of the FI coefficients, the vacuum in the Coulomb branch breaks partially supersymmetry. By making a different choice, we will show that supersymmetry is fully broken.

Using the ansatz (\ref{eqn:partprep}) and restricting the model to imaginary $e_a$, the stationarity conditions (\ref{eqn:statcoul}) reduce to:
\begin{equation}
\label{eqn:redstatcoul}
\begin{split}
&\langle f_1''\rangle=ie_1\kappa_1\pm \sqrt2i\xi_1\kappa_1,\\
&\langle f_2''\rangle=ie_2\kappa_2\pm \sqrt2i\xi_2\kappa_2.
\end{split}
\end{equation}
These equations can generically be solved for given functions $f_a\left(X^a\right)$. The sign ambiguity in the imaginary part of these VEVs can be set by demanding that the special-K\"ahler metric $h_{ab}$ be positive-definite. In \cite{Itoyama:2007kk}, the sign of all terms $\xi_a\kappa_a$ was chosen to be the same. Here we impose that the metric is positive-definite for opposite signs: $(\xi_1\kappa_1)(\xi_2\kappa_2)<0$. Without loss of generality, we take the $+$ sign in the first line.

 Thus the VEVs of the auxiliary fields become:
 \begin{equation}
 \label{eqn:auxvevstm}
 \begin{split}
 &\langle F^{x^1}\rangle=-\frac{1}{4\kappa_1} \textrm{ , } \langle F^{x^2}\rangle=\frac{g_1}{4g_2\kappa_1},\\
 &\langle D^1\rangle=-\frac{1}{2\sqrt2\kappa_1} \textrm{ , } \langle D^2\rangle=-\frac{g_1}{2\sqrt2g_2\kappa_1},
 \end{split}
 \end{equation}
 where we took into account the constraint (\ref{eqn:addconstr}). The VEV of the auxiliary fields $F^{q^u}$ vanishes in the Coulomb phase. Inserting these results into the expressions of the supersymmetry breaking scales (\ref{eqn:susyscales}) gives
 \begin{equation}
 \label{eqn:sscalestm}
 \begin{split}
 &\left(\Lambda^1_{SUSY}\right)^2=\frac{g_1\xi_2}{2Ng_2\kappa_1},\\
 &\left(\Lambda^2_{SUSY}\right)^2=\frac{\xi_1}{2N\kappa_1}=\frac{g_2\xi_1}{g_1\xi_2}\left(\Lambda^1_{SUSY}\right)^2.
 \end{split}
 \end{equation}
 As discussed above the difference between the two scales is proportional to the deformation $1/\kappa_1$. In order for them to be well separated, one has to require for example: $g_2\xi_1\gg g_1\xi_2$. Thereafter we will assume\footnote{Notice that we could also play with the relative strength of the two gauge couplings.} that $\xi_1$ is much bigger than $\xi_2$ and the two gauge couplings are both either strong or weak.
 
 One may expect that the system behaves at the intermediate scale $\Lambda^1_{SUSY}\ll \Lambda_{int}\ll \Lambda^2_{SUSY}$ like in a phase breaking partially supersymmetry. Indeed, the two goldstini in the limit $\xi_1\gg\xi_2$ become:
 \begin{equation}
 \label{eqn:goldstm}
 \begin{split}
&\eta_1=\frac{1}{2\sqrt2N}\left[-\xi_1\left(\psi^{x^1}-i\lambda^1\right)+\xi_2\left(\psi^{x^2}+i\lambda^2\right)\right],\\
&\eta_2=\frac{1}{2\sqrt2N}\left[i\xi_1\left(\psi^{x^1}-i\lambda^1\right)+i\xi_2\left(\psi^{x^2}+i\lambda^2\right)\right].
\end{split}
\end{equation}
In this limit the term proportional to $\xi_2$ can be neglected, leading to the same expression up to a phase. Thus, at an intermediate scale, the two goldstini effectively coincide.

One could also consider the mass spectrum in this limit to see whether and how fields are organized in $\mathcal N=1$ multiplets. The scalar and fermion masses are derived from the corresponding mass matrices (\ref{eqn:scalmass}) and (\ref{eqn:fermmass}), respectively. One has to take also into account the normalization of the gauge kinetic terms which involve the nontrivial metric $h_{ab}$. Finally the gauge bosons $A^a_{\mu}$ are massless in the Coulomb phase. The full mass spectrum is summarized in the table below:
\begin{center}
\begin{tabular}{| c | c || c | c |}
\hline
\multicolumn{2}{| c ||}{Exact expression} & \multicolumn{2}{c |}{Limit $\xi_1\gg\xi_2$}\\
\hline
Field & Square mass & Field & Square mass\\
\hline\hline
$q^1$ & $\left|m_{eff}\right|^2-\frac{g_1}{\sqrt2\kappa_1}$ & $q^1$ & $\left|m_{eff}\right|^2$\\
$q^2$ & $\left|m_{eff}\right|^2+\frac{g_1}{\sqrt2\kappa_1}$  & $q^2$ & $\left|m_{eff}\right|^2$\\
$x^1$ & $\frac{\left|f_1'''\right|^2}{32\xi_1^2\kappa_1^4}$ & $x^1$ & $\frac{\left|f_1'''\right|^2}{32\xi_1^2\kappa_1^4}$\\
$x^2$ & $\frac{g_1^4\left|f_2'''\right|^2}{32g_2^4\xi_2^2\kappa_1^4}$ & $x^2$ & $\frac{g_1^4\left|f_2'''\right|^2}{32g_2^4\xi_2^2\kappa_1^4}$ \\
\hline
$\chi^1$ & $\left|m_{eff}\right|^2$ & $\chi^1$ & $\left|m_{eff}\right|^2$\\
$\chi^2$ & $\left|m_{eff}\right|^2$ & $\chi^2$ & $\left|m_{eff}\right|^2$\\
$\frac{\psi^{x^1}+i\lambda^1}{\sqrt2}$ & $\frac{\left|f_1'''\right|^2}{32\xi_1^2\kappa_1^4}$ & $\frac{\psi^{x^1}+i\lambda^1}{\sqrt2}$ & $\frac{\left|f_1'''\right|^2}{32\xi_1^2\kappa_1^4}$\\
$\frac{\psi^{x^2}-i\lambda^2}{\sqrt2}$ & $\frac{g_1^4\left|f_2'''\right|^2}{32g_2^4\xi_2^2\kappa_1^4}$ & $\frac{\psi^{x^2}-i\lambda^2}{\sqrt2}$ & $\frac{g_1^4\left|f_2'''\right|^2}{32g_2^4\xi_2^2\kappa_1^4}$\\
$\eta_1$ & 0 & $\frac{\psi^{x^1}-i\lambda^1}{\sqrt2}$ & 0\\
$\eta_2$ & 0 & $\frac{\psi^{x^2}+i\lambda^2}{\sqrt2}$ & 0\\
\hline
$A^1_{\mu}$ & 0 & $A^1_{\mu}$ & 0\\
$A^2_{\mu}$ & 0 & $A^2_{\mu}$ & 0\\
\hline
\end{tabular}
\end{center}
where $m_{eff}$ is defined in (\ref{eqn:meff}).

The mass splitting of the scalars $q^u$ with respect to the hyperini $\chi^u$ confirms again that supersymmetry is fully broken on the vacuum. This Coulomb phase vacuum is stable only if $|m_{eff}|^2 > |\frac{g_1}{\sqrt2\kappa_1}|$. It turns out that $|\frac{g_1}{\kappa_1}|$ must be small compared to $|m_{eff}|^2$ if one wants to recover an approximate $\mathcal N=1$ mass spectrum at the intermediate energy scale $\Lambda_{int}$. The spectrum in this limit is displayed in the right part of the table above, where the two goldstino expressions coincide following (\ref{eqn:goldstm}).

The matter fields can be embedded in $\mathcal N=1$ multiplets with respect to any combination of the two supersymmetries\footnote{We showed in section \ref{susybs} that partial SUSY breaking can possibly preserve only the combinations $\mathcal Q\pm i\mathcal{\tilde Q}$ for the current choice of FI parameters.}. All gauge fields associated to the first ($a=1$) $U(1)$ factor can be arranged in $\mathcal N=1$ multiplets of the preserved supersymmetry $\mathcal Q+i\mathcal{\tilde Q}$, while gauge fields associated to the second $U(1)$ ($a=2$) are organized in $\mathcal N=1$ multiplets of the supersymmetry $\mathcal Q-i\mathcal{\tilde Q}$. Due to the separability of the prepotential (\ref{eqn:partprep}), the two gauge sectors are decoupled and the model indeed possesses an effective $\mathcal N=1$ supersymmetry at the intermediate scale $\Lambda_{int}$.

Notice that the separable prepotential ({\ref{eqn:partprep}) can be replaced by a more general function by introducing a coupling between the superfields $X^1$ and $X^2$. As long as this coupling/deformation of the previous model remains small and can be treated perturbatively, the vacuum described above will remain stable, since all the scalar masses are strictly positive.
\section{Comments on light fermions and pseudo-reality in $\mathcal N=2$ supersymmetric theories\label{chir}}
$\mathcal N=2$ theories exhibit many compelling features. The most impressive is probably the possibility to derive the exact non-perturbative dynamics of gauge models \cite{Seiberg:1994rs,Seiberg:1994aj}. Another interesting property of these theories is that the superpotential is tightly constrained in comparison with $\mathcal N=1$ supersymmetry where any holomorphic function is compatible with SUSY. For example Yukawa and gauge couplings are unified while one can only write FI terms in the gauge sector.

In spite of these advantages, $\mathcal N=2$ theories suffer the lack of chirality of their particle spectrum. Chirality manifests itself in the Standard Model with the property of quarks and leptons to belong to complex, not self-conjugate representations of the gauge group $SU(3)_c\times SU(2)_L\times U(1)_Y$. In a $\mathcal N=2$ extension of the SM, quarks and leptons cannot be part of vector multiplets, since the adjoint representation is real. They are therefore embedded in hypermultiplets. In addition to a scalar partner, which is reminiscent from the first supersymmetry, each particle and sparticle has a "mirror" partner, which is in the conjugate representation. Any phenomenologically viable $\mathcal N=2$ extension of the SM must include a mechanism for recovering chirality. Different approaches have been proposed in the literature. In the context of string theory, chirality can emerge by appropriately choosing the compactification to a four-dimensional theory \cite{Candelas:1985en}. The second supersymmetry is generally broken at the compactification scale, although it may survive (at lowest order) in a subsector of the theory involving for instance gauge multiplets. A purely field theoretic mechanism consists in adding explicit SUSY breaking mass terms for the ``mirror" fermions, in order to lift the degeneracy with SM particles \cite{delAguila:1984qs,Girardello:1997hf,Polonsky:2000zt}. These necessarily should arise via Yukawa couplings with the Higgs field breaking hardly $\mathcal N=2$. Moreover, in order to make the ``mirror" fermions much heavier than their SM partners one is  generally facing a strong coupling problem since the corresponding Yukawa couplings become large.

These two approaches do not deal with truly $\mathcal N=2$ supersymmetric models. In the former, supersymmetry is broken at the compactification scale, making the interpretation in terms of $\mathcal N=2$ multiplets difficult, while the second mechanism breaks also explicitly $\mathcal N=2$. Here, we want to stress one feature of hypermultiplets that may help to lift the degeneracy between SM fermions and mirrors, while keeping $\mathcal N=2$ invariance. When hypermultiplets belong to a pseudo-real representation of the gauge group, they also form a CPT self-conjugate representation \cite{Breitenlohner:1981sm,Mezincescu:1983rn,Derendinger:1984bu} without the need for doubling the spectrum, i.e., in this case it has been shown that the following reality condition\footnote{$A^i_{\;a}$ are the scalars and $\chi_a,\psi^a$ the fermions of the hypermultiplet. This notation stresses that the scalars are doublets and hyperini are singlets of the $SU(2)_R$ symmetry.}
\begin{equation}
\label{eqn:realcondprr}
\bar A^{i}_{\;a}=\epsilon^{ij}C_{ab}A_{j}^{\;b} \textrm{\quad , \quad} \bar\chi^{\dot\alpha a}=\epsilon^{\dot\alpha\dot\beta}C^{ab}\bar\psi_{\dot\beta b}
\end{equation}
can be imposed, implying again among other things that the number of real scalars is a multiple of four. $C_{ab}$ is the antisymmetric invariant metric of the pseudo-real representation\footnote{For the doublet of $SU(2)$, this is simply the antisymmetric invariant tensor $\epsilon_{ab}$.}. It is crucial for consistency of these equations that $C_{ab}$ be antisymmetric. This formulation of the hypermultiplet finally possesses the same field content as two $\mathcal N=1$ chiral superfields. The property (\ref{eqn:realcondprr}) does not allow to eliminate the mirror particles, but is nevertheless interesting as we discuss now.

When hypermultiplets are in real or complex representations, gauge invariant fermion mass terms are allowed. The expression $\chi^{\alpha}_{\;a}\psi_{\alpha}^{\;a}$ is indeed gauge invariant. However, when the condition (\ref{eqn:realcondprr}) is satisfied, this quantity vanishes. This is a particular realization of a more general feature of pseudo-real representations, which states that their quadratic invariants are antisymmetric \cite{Derendinger:1984bu}. In other words, the singlet that appears in the tensor product of two pseudo-real representations $R$
\begin{equation}
\label{eqn:pseudorealte}
R\otimes R =1_A\oplus\ldots
\end{equation}
corresponds to the antisymmetric tensor $C_{ab}$. The immediate consequence of this property is that a $\mathcal N=2$ supersymmetric mass term can only be written for two different hypermultiplets. In a model with an odd number of multiplets, at least one is necessarily massless.

The occurrence of massless fermions in addition to the goldstini suggests a possibility of readdressing the problem of chirality towards a phenomenologically viable $\mathcal N=2$ theory. In the presence of naturally light fermions, the problem of chirality in $\mathcal N=2$ supersymmetry with pseudo-real representations consists in finding some $\mathcal N=2$ mechanism that selects quarks and leptons as these light states while making mirrors heavy. Keeping this in mind, it would be interesting to study whether a supersymmetric Grand Unified Theory can be built with pseudo-real representations. The simplest model deals with doublets of a $SU(2)$ gauge group. In another non-trivial example\footnote{A list of some pseudo-real representations is provided for example in \cite{Derendinger:1984bu}.}, hypermultiplets form a threefold antisymmetric tensor of $SU(6)$. This is a twenty-dimensional representation. However, albeit this group has $SU(5)$ as a subgroup, the $\textbf{20}$ of $SU(6)$ is not directly relevant for chiral model-building.
\section{Conclusion\label{conc}}
In this work we performed the analysis of the minimal ingredients required for global $\mathcal N=2$ supersymmetry to be sequentially spontaneously broken at two distinct scales. In order to evade the no-go theorem which ensures that all the supersymmetries are simultaneously broken, one has to introduce a deformation $\frac{1}{\kappa_a}$ in the realization of the second supersymmetry. This deformation induces magnetic FI terms and allows partial SUSY breaking. Several difficulties were pointed out in the generalization of this mechanism to theories with charged matter. To keep $\mathcal N=2$ invariance under control, we chose to express the action of our model in terms of chiral $\mathcal N=2$ superfields. While this is a standard way to describe Maxwell multiplets or their non-abelian generalization, an off-shell description of hypermultiplets requires the more involved harmonic superspace. One can however avoid it by working with single-tensor multiplets, which are dual to hypermultiplets with associated $U(1)$ (shift) isometries. Invariance under the deformed supersymmetry then introduces a term which linearly depends on a non-dynamical four-form field $C_{\mu\nu\rho\sigma}$. Consistency of its equation of motion imposes a constraint on the coefficient it is multiplied by, which is proportional to the magnetic FI parameters $\frac{1}{\kappa_a}$. This constraint can only be non-trivially satisfied if the hypermultiplets are charged under at least two $U(1)$'s. We then focused on the minimal model formed by one hypermultiplet with a canonical kinetic term, and two Maxwell multiplets with an arbitrary holomorphic prepotential.

We determined in this context the vacuum structure of our model. While the Coulomb phase can only be analyzed in the hypermultiplet formalism, the single-tensor formalism can also be used to describe off-shell the Higgs phase. The study of possible vacua simplified the computation of the two supersymmetry-breaking scales that were derived from the supersymmetric variations of the two goldstini $\delta_i\eta_j$. As expected, according to the no-go theorem above, we found that their degeneracy is lifted only in the presence of a deformation. We discussed a particular toy example involving just two $U(1)$'s, one hypermultiplet charged under both of them, and a separable but non-trivial effective prepotential. By fine-tuning the FI coefficients, we showed that the two scales can be freely adjusted. We then analyzed the spectrum in an intermediate regime and found that the system behaves like in a phase of partially broken SUSY.

One interesting consequence of the breaking of global $\mathcal N=2$ supersymmetry in two steps is to help to reconcile \lq\lq bottom-up" and \lq\lq top-down" approaches to high energy physics. In the former, starting from the SM, $\mathcal N=1$ SUSY, if it is to be found at the LHC, provides an elegant and efficient solution to the problem of stabilization of the electroweak scale. In the latter, effective four-dimensional models originating from string theory often exhibit extended supersymmetry, at least in the gauge sector of the theory. If the two scales are of different orders of magnitude, the two behaviors are present. A drawback of the mechanism discussed in this work is the need for introducing by ``hand" the electric and magnetic parameters. One may then try to generate these FI terms quantum mechanically instead of putting them in by hand (the resulting dynamical completion cannot be pure $\mathcal N=2$ Super Yang-Mills or another gauge theory with a linear anomaly multiplet \cite{Antoniadis:2010nj}; it would also be subject to the restrictions of \cite{Komargodski:2009pc}). One interesting avenue would be to precisely relate the two SUSY-breaking scales computed in section \ref{gold} to the gravitini masses in supergravity models. One could then analyze the restrictions imposed by string theory on dynamically-generated field-dependent FI terms \cite{Dine:1987xk} in the four-dimensional effective $\mathcal N=2$ supergravity (SUGRA) theory.

At this point we are far from being able to assert that the SM can be extended to a viable $\mathcal N=2$ theory. It would be nevertheless worthwhile to put together the different features that we pointed out or reviewed in this note: supersymmetry breaking at two distinct scales and the presence of massless (matter) fermions in gauge theories involving pseudo-real representations to see if the situation improves.

\section*{Acknowledgements}
We are grateful to Matthew Buican for stimulating discussions and for comments on the manuscript. We also wish to thank Sergio Ferrara, Riccardo Rattazzi and Claudio Scrucca for useful conversations. This work was supported in part by the European Commission under the ERC Advanced Grant 226371 and the contract PITN-GA-2009-237920 and by the Swiss Science Foundation.
\appendix
\section{$\mathbf{\mathcal N=2}$ supersymmetry in the hypermultiplet formalism\label{app:varhyp}}
In this paper we proved the invariance of the Lagrangian (\ref{eqn:lagtothyp}) under $\mathcal N=2$ supersymmetry by first making this symmetry manifest using $\mathcal N=2$ superspace, and then by performing a duality transformation from the single-tensor formalism to the hypermultiplet one. One could straight verify its invariance by working with hypermultiplets in $\mathcal N=1$ superspace from the very beginning. The (linearly realized) $\mathcal N=1$ supersymmetry is then manifest, but the second one must be carefully checked. This approach was pioneered in \cite{Hull:1985pq} and more recently generalized in \cite{Jacot:2010df}.

For convenience we repeat the Lagrangian of our model:
\begin{equation}
\label{eqn:rlagtothyp}
\begin{split}
\mathcal{L}&=\int d^4\theta \left[\bar{Q}^1 e^{-2g_aV^a}Q^1 + \bar{Q}^2 e^{2g_aV^a}Q^2+\frac{i}{2}\left(\bar{\mathcal{F}}_{a}X^a-\mathcal{F}_a\bar X^a\right)
+\xi_aV^a\right]\\
&+\int d^2\theta\left[-\frac{i}{4}\mathcal{F}_{ab}W^{a}W^{b}+\left(m+\sqrt2i g_aX^a\right)Q^1Q^2-\frac{e_a}{4}X^a-\frac{i}{4\kappa_a}\mathcal F_a\right]+ \textrm{ h.c.}\\
\end{split}
\end{equation}
As explained in detail the constraint (\ref{eqn:addconstr}) is crucial for supersymmetry invariance. It has already been applied to (\ref{eqn:rlagtothyp}).

The transformation laws of the gauge and matter fields with respect to the first supersymmetry are \cite{Wess:1992cp}:
\begin{equation}
\label{eqn:compvarn1}
\begin{array}{ll}
\delta q^u=\sqrt2\epsilon\chi^u\textrm{,} & \delta\chi^u=\sqrt2 F^{q^u}+\sqrt2iD\!\!\!\!/\, q^u\bar\epsilon\textrm{ ,}\\
\delta x^a=\sqrt2\epsilon\psi^{x^a}\textrm{,} & \delta \psi^{x^a}=\sqrt2 F^{x^a}+\sqrt2i\partial\!\!\!/\, x^a\bar\epsilon\textrm{ ,}\\
\delta A^a_{\mu}=i\epsilon\sigma_{\mu}\bar\lambda^a-i\lambda^a\sigma_{\mu}\bar\epsilon\textrm{ ,} & \delta\lambda^a=i\epsilon D^a+\sigma^{\mu\nu}\epsilon F^a_{\mu\nu}\textrm{ ,}
\end{array}
\end{equation}
where the covariant derivative is given by $D_{\mu}q^u=\partial_{\mu}q^u+A^a_{\mu}k_a$, $k_a=g_ak$ and $k$ is the  triholomorphic Killing vector field of the $U(1)$ isometry of the hyper-K\"ahler manifold \cite{Hull:1985pq}: $k^u=-i(q^1,-q^2)$. The full expression of the auxiliary fields reads:
\begin{equation}
\label{eqn:auxfwf}
\begin{split}
&F^{q^u}=i\bar\Omega^u_{\;\;\bar v}\left(m-\sqrt2ig_a\bar x^a\right)\bar k^{\bar v},\\
&F^{x^a}=h^{ab}\left(\sqrt2ig_b\bar q^{\bar1}\bar q^{\bar2}+\frac{\bar e_b}{4}-\frac{i\textrm{Re}\mathcal{F}_{bc}}{4\kappa_c}\right)-\frac{1}{4\kappa_a}+\frac{i}{4}h^{ab}\bar{\mathcal F}_{bcd}\bar\lambda^c\bar\lambda^d,\\
&D^a=-\frac{1}{2}h^{ab}\left[-2g_b\left(\left|q^1\right|^2-\left|q^2\right|^2\right)+\xi_b\right]-\frac{1}{2\sqrt2}h^{ab}\mathcal F_{bcd}\psi^{x^c}\lambda^d,
\end{split}
\end{equation}
where $\Omega_{uv}=\epsilon_{uv}$ is the 2-form defining the three complex structures of the hyper-K\"ahler manifold \cite{Hull:1985pq,DeJaegher:1997ka}.

The second supersymmetry variations mix in this formalism different $\mathcal N=1$ superfields. For the gauge fields they were given by (\ref{eqn:varmaxdef}), while for the hypermultiplet they become
\begin{equation}
\label{eqn:hypvarn2}
\tilde\delta Q^u=-\frac{1}{2}\bar\Omega^{uv}\bar D^2\left[\left(\bar Q_v+2ig_a\bar k_vV^a\right)\bar{\tilde\epsilon}\bar\theta\right]-2imk^u\tilde\epsilon\theta
\end{equation}
in the Wess-Zumino gauge.

The field expansion of these superfield transformation laws is given by:
\begin{equation}
\label{eqn:compvarn2}
\begin{array}{ll}
\tilde\delta q^u=-\sqrt2\bar\Omega^{u}_{\;\;\bar v}\bar{\tilde\epsilon}\bar\chi^{\bar v}\textrm{,} & \tilde\delta\chi^u=\sqrt2\tilde\epsilon\tilde F^{q^u}+\sqrt2i\bar\Omega^{u}_{\;\;\bar v}D\!\!\!\!/\, \bar q^{\bar v}\bar {\tilde\epsilon}\textrm{ ,}\\
\tilde\delta x^a=\sqrt2\tilde\epsilon\lambda^a\textrm{,} & \tilde\delta\psi^{x^a}=i\tilde\epsilon D^a+\sigma^{\mu\nu}\tilde\epsilon F^a_{\mu\nu}\textrm{ ,}\\
\tilde\delta A^a_{\mu}=-i\tilde\epsilon\sigma_{\mu}\bar\psi^{\bar x^a}+i\psi^{x^a}\sigma_{\mu}\bar{\tilde\epsilon}\textrm{ ,} & \tilde\delta\lambda^a=\sqrt2\tilde\epsilon\bar F^{\bar x^a}+\frac{1}{\sqrt2\kappa_a}\tilde\epsilon+\sqrt2i\partial\!\!\!/\, x^a\bar{\tilde\epsilon}.
\end{array}
\end{equation}
The only new quantity compared with (\ref{eqn:compvarn1}) is $\tilde F^{q^u}$. It is defined as
\begin{equation}
\label{eqn:auxft}
\tilde F^{q^u}=-i\left(m-\sqrt2ig_a\bar x^a\right)k^u.
\end{equation}
One can verify that these variations satisfy the supersymmetry algebra:
\begin{equation}
\label{eqn:alghyp2}
[\tilde\delta_1,\tilde\delta_2]Q^u=-2i\left(\tilde\epsilon_1\sigma^{\mu}\bar{\tilde\epsilon}_2-\tilde\epsilon_2\sigma^{\mu}\bar{\tilde\epsilon}_1\right)\partial_{\mu}Q^u.
\end{equation}
However, it closes only on-shell. Finally the constraint (\ref{eqn:addconstr}) could be derived by imposing the invariance of the model (\ref{eqn:rlagtothyp}) under the transformations (\ref{eqn:varmaxdef}) and (\ref{eqn:hypvarn2}).

To determine whether partial or complete supersymmetry breaking occurs, an efficient method consists in analyzing the mass spectrum. The scalar mass terms of the model (\ref{eqn:rlagtothyp}) are obtained by taking the second derivatives of the scalar potential $V_S$ (\ref{eqn:scalpothyp}). Using the stationarity conditions the components of the scalar mass matrix\footnote{ $I,J$ describe both holomorphic and antiholomorphic indices.} $\left(M^2_0\right)_{IJ}\phi^I\phi^{J}$ are found to be given by:
\begin{equation}
\label{eqn:scalmass}
\begin{split}
&\left(M^2_0\right)_{q^1\bar q^{\bar1}}=\left|m_{eff}\right|^2+g_ah^{ab}g_b\left(\left|q^1\right|^2+2\left|q^2\right|^2\right)+g_aD^a,\\
&\left(M^2_0\right)_{q^1\bar q^{\bar2}}=g_ah^{ab}g_b\bar q^{\bar1}q^2,\\
&\left(M^2_0\right)_{q^2\bar q^{\bar2}}=\left|m_{eff}\right|^2+g_ah^{ab}g_b\left(2\left|q^1\right|^2+\left|q^2\right|^2\right)-g_aD^a,\\
&\left(M^2_0\right)_{q^1\bar x^a}=-\sqrt2ig_am_{eff}\bar q^{\bar1}-\frac{1}{\sqrt2}\bar{\mathcal F}_{abc}h^{bd}g_d\left[\left(F^{x^c}+\frac{1}{2\kappa_c}\right)q^2+\frac{i}{\sqrt2}D^c\bar q^{\bar1}\right],\\
&\left(M^2_0\right)_{q^2\bar x^a}=-\sqrt2ig_am_{eff}\bar q^{\bar2}-\frac{1}{\sqrt2}\bar{\mathcal F}_{abc}h^{bd}g_d\left[\left(F^{x^c}+\frac{1}{2\kappa_c}\right)q^1-\frac{i}{\sqrt2}D^c\bar q^{\bar2}\right],\\
&\begin{split}\left(M^2_0\right)_{x^a\bar x^b}&=\frac{1}{4}\mathcal F_{ace}h^{cd}\bar{\mathcal F}_{bdf}\left[F^{x^e}\bar F^{\bar x^f}+\left(F^{x^e}+\frac{1}{2\kappa_e}\right)\left(\bar F^{\bar x^f}+\frac{1}{2\kappa_f}\right)+D^eD^f\right]\\
&+2g_ag_b\left(\left|q^1\right|^2+\left|q^2\right|^2\right),\end{split}\\
&\left(M^2_0\right)_{q^1q^1}=g_ah^{ab}g_b\left(\bar q^{\bar1}\right)^2,\\
&\left(M^2_0\right)_{q_1q_2}=-\sqrt2ig_aF^{x^a}-g_ah^{ab}g_b\bar q^{\bar1}\bar q^{\bar2},\\
&\left(M^2_0\right)_{q^2q^2}=g_ah^{ab}g_b\left(\bar q^{\bar2}\right)^2,\\
&\left(M^2_0\right)_{q^1x^a}=\sqrt2ig_a\bar{m}_{eff}\bar q^{\bar1}+\frac{1}{\sqrt2}\mathcal F_{abc}h^{bd}g_d\left(F^{x^c}q^2+\frac{i}{\sqrt2}D^c\bar q^{\bar1}\right),\\
&\left(M^2_0\right)_{q^2x^a}=\sqrt2ig_a\bar{m}_{eff}\bar q^{\bar2}+\frac{1}{\sqrt2}\mathcal F_{abc}h^{bd}g_d\left(F^{x^c}q^1-\frac{i}{\sqrt2}D^c\bar q^{\bar2}\right),\\
&\begin{split}\left(M^2_0\right)_{x^ax^b}&=-\frac{1}{4}\mathcal F_{ace}h^{cd}\mathcal F_{bdf}\left[F^{x^e}\left(\bar F^{\bar x^f}+\frac{1}{2\kappa_f}\right)+F^{x^f}\left(\bar F^{\bar x^e}+\frac{1}{2\kappa_e}\right)+D^eD^f\right]\\
&+\frac{i}{2}\mathcal F_{abcd}\left[F^{x^c}\left(\bar F^{\bar x^d}+\frac{1}{2\kappa_d}\right)+\frac{1}{2}D^cD^d\right].\end{split}
\end{split}
\end{equation}
The remaining components are simply obtained by complex conjugation of the previous ones. The fermion mass matrix $\frac{1}{2}\left(M_{1/2}\right)_{ij}\psi^i\psi^j$ can be however immediately deduced from the field expansion of (\ref{eqn:rlagtothyp}):
\begin{equation}
\label{eqn:fermmass}
\begin{split}
&\left(M_{1/2}\right)_{\chi^1\chi^1}=\left(M_{1/2}\right)_{\chi^2\chi^2}=0\, ,\\
&\left(M_{1/2}\right)_{\chi^1\chi^2}=m_{eff},\\
&\left(M_{1/2}\right)_{\chi^1\psi^{x^a}}=\sqrt2ig_aq^2 \textrm{ , } \left(M_{1/2}\right)_{\chi^2\psi^{x^a}}=\sqrt2ig_aq^1,\\
&\left(M_{1/2}\right)_{\chi^1\lambda^a}=\sqrt2ig_a\bar q^{\bar1} \textrm{ , } \left(M_{1/2}\right)_{\chi^2\lambda^a}=-\sqrt2ig_a\bar q^{\bar2},\\
&\left(M_{1/2}\right)_{\psi^{x^a}\psi^{x^b}}=-\frac{i}{2}\mathcal F_{abc}\left(\bar F^{\bar x^c}+\frac{1}{2\kappa_c}\right),\\
&\left(M_{1/2}\right)_{\psi^{x^a}\lambda^b}=-\frac{1}{2\sqrt2}\mathcal F_{abc}D^c,\\
&\left(M_{1/2}\right)_{\lambda^a\lambda^b}=-\frac{i}{2}\mathcal F_{abc}F^{x^c}.
\end{split}
\end{equation}

\bibliographystyle{unsrturl}
\bibliography{biblio}
\end{document}